# CAST RESULTS AND AXION REVIEW

T. GERALIS
*Institute of Nuclear Physics, NCSR Demokritos*
*GR-15310 AG. PARASKEVI, ATHENS, GREECE*
*For the CAST Collaboration*

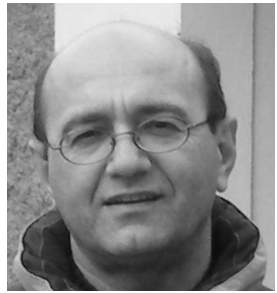

We present results from the CERN Axion Solar Telescope (CAST) and the Axion Dark Matter eXperiment (ADMX), together with a brief review on prospects on Axion searches with a variety of experimental techniques. CAST has explored masses up to 0.64 eV setting the most stringent limit on the axion-photon coupling, apart for the μeV region where ADMX is the most competitive experiment. CAST is aiming at surpassing the 1eV WMAP upper limit and possibly revisiting the operation in vacuum with extra sensitive X-ray detectors, while ADMX, using improved extra sensitive SQUID amplifiers will explore the μeV mass range.

**CAST Collaboration**
S.Aune, K.Barth, A.Belov, S.Borghi, H.Brauninger, G.Cantatore, J.M.Carmona, S.A.Cetin, J.I.Collar, T.Dafni, M.Davenport, L. Di Lella, C.Eleftheriadis, N.Elias, C.Ezer, G.Fanourakis, E.Ferrer-Ribas, H.Fischer, J.Franz, P.Friedrich, J.Galan, E.Gazis, T.Geralis, I.Giomataris, S.Gninenko, H.Gomez, R.Hartmann, F.Haug, M.Hasinoff, D.H.H.Hoffmann, F.J.Iguaz, I.G.Irastorza, J.Jacoby, K.Jakovcic, D.Kang, T.Karageorgopoulou, M.Karuza, K.Koenigsmann, R.Kotthaus, M.Krcmar, K.Kousouris, M.Kuster, B.Lakic, A.Liolios, A.Ljubivcic, V.Lozza, G.Lutz, G.Luzon, D.Miller, A.Mirizzi, J.Morales, H.Mota, T.Niinikoski, A.Nordt, T.Papaevangelou, M.Pivovaroff, G.Raiteri, G.Raffelt, H.Riege, A.Rodriguez, J.Ruz, I.Savvidis, Y.Semertzidis, P.Serpico, P.S.Silva, S.K.Solanki, R.Soufli, L.Stewart, M.Tsagri, K. van Bibber, T.Vafeiadis, J.Villar, J.Vogel, L.Walckiers, Y.Wong, K.Zioutas.

## 1 Introduction

Quantum Chromo Dynamics (QCD) is the theory describing the interaction of particles carrying strong charge, conventionally named color that exists in three kinds: red, green and blue. QCD is described by the SU(3) algebra and is invariant under Lorentz and local gauge transformations. QCD doesn't respect the chiral symmetry unless we are limited in the light quarks and we consider them as massless. The non-abelian character of the SU(3) symmetry group leads to a Lagrangian in which the Gauge fields interact because they carry color themselves. This property introduces complex non linear phenomena which result in a non trivial topological structure of the vacuum. Instanton solutions consist of tunneling effects between non equivalent topological classes. The true vacuum is actually a linear combination of an infinite number of degenerate vacua, denoted by |n>, the so called θ-vacuum:

$$|\theta\rangle = \sum_{n=-\infty}^{\infty} e^{in\theta} |n\rangle$$

The parameter θ is an observable and introduces in the Lagrangian a CP-Violating term. Since we have already a CP-Violating term from the EW quark mass mixing, the term including both CP-Violating contributions is:

$$\mathcal{L}_\theta = \frac{g^2 \bar{\theta}}{32\pi^2} G^\alpha_{\mu\nu} \tilde{G}^{\alpha\mu\nu} \quad \text{with} \quad \bar{\theta} = \theta + Arg(\det M)$$

Theoretical calculations for the neutron Electric Dipole Moment give:

$$d_n = \frac{e}{m_n} \bar{\theta} \frac{m_u m_d}{m_u + m_d} \frac{1}{\Lambda_{QCD}} \approx 10^{-16} \bar{\theta} e \cdot cm$$

From the current experimental limit: $d_n < 2.9 \times 10^{-26} e \cdot cm$ [1] we conclude that $\bar{\theta} \approx 10^{-10}$ and the question is why those contributions coming from two deferent sectors cancel each other to such a high precision (Strong CP-problem).

R. Peccei and H. Quinn in 1977 invented a model that was later complemented by Weinberg and Wilczek [2], that introduced a new global symmetry U(1)$_{PQ}$, the spontaneous braking of which offered the cancelation to the θ – term:

$$\mathcal{L}_a = \left( \bar{\theta} - \frac{a(x)}{f_a} \right) \frac{1}{f_a} \frac{g}{8\pi} G_a^{\mu\nu} \tilde{G}_{a\mu\nu}$$

At the same time it introduced a neutral pseudoscalar Nambu – Goldstone (NG) boson, the Axion. The scale of the symmetry breaking was thought to be at the EW level, but this was quickly ruled out experimentally since no axion was found with the predicted mass of the order of ~1 MeV.

Later models, DFSZ (Dine, Fischler, Srednicki, Zhitnisky) [3] and KSVZ (Kim, Shifman, Vainstein, Zakharov) [4] pushed the breaking scale much higher making thus the coupling extremely small (invisible axion). Those models introduced either new Higgs fields or new exotic quarks with the KSVZ not allowing axion interactions to leptons justifying the term QCD axions.

Axion, as every NG boson couples derivatively to matter. The axion is special because it couples to two gluons with a non derivative term and at low energies this term generates the axion mass:

$$m_a = \frac{f_\pi m_\pi}{f_a} \frac{\sqrt{m_u m_d}}{m_u + m_d} = 0.6 eV \left(\frac{10^7 GeV}{f_a}\right)$$

All effective coupling constants of axions with matter and radiation depend on the inverse of the symmetry breaking scale and are therefore linear with axion mass. In particular for the axion to photon coupling:

$$g_{a\gamma} = \frac{\alpha}{2\pi}\left(\frac{E}{N} - \frac{2}{3}\frac{4+z}{1+z}\right)\frac{1+z}{z^{1/2}}\frac{1}{m_\pi f_\pi} m_a$$

E/N is model dependent constant of the order 1, and $z = m_u/m_d$.

Axion thus appears to have very interesting properties, it is a neutral pseudoscalar and it has very low mass and very small coupling. Those properties make it a suitable candidate for Cold Dark Matter (CDM) if its mass lies at the μeV range and for Hot Dark Matter (HDM) if it possesses higher mass. Its contribution to Ω is: $\Omega_a \approx (5\mu eV/m_a)^{7/6}$. Axions couple to photons via triangular loops. The interaction term in the Lagrangian is: $\mathcal{L}_{int} = g_{a\gamma\gamma} a \vec{E}\cdot\vec{B}$. The Axion production and detection are based on the so called Primakoff conversion similar to the $\pi^0$ conversion.

**2 Axion experimental techniques**

The axion discovery may offer two solutions at the same time for two long standing problems, the Strong CP problem and the Dark Matter problem. Experimental effort to detect the invisible axion may be classified in different techniques like Laser induced axions, Telescope searches, Bragg diffraction, Geomagnetic axion conversion, Helioscope searches and Microwave cavity searches. Laser induced axions exploit the passage of laser beams through very strong magnetic fields to generate the axions and reconvert it back to light in another magnetic field behind a wall, the so called: light shine through the wall experiments [5]. The full control of the generation and the conversion of axions is an advantage of those experiments but on the other hand the probability to produce axions goes with the forth power of the coupling constant. GammeV [6] is a "light shine through the wall" experiment especially set to investigate the PVLAS [7] experiment signal. OSQAR[8] and ALPS [9] are experiments that are under construction and they use in series LHC and DESY magnets correspondingly. Their sensitivity may reach $g_{a\gamma\gamma} \sim 10^{-11} \text{GeV}^{-1}$. Telescope searches [10] may explore masses $m_a < 10$ μeV. Bragg diffraction in crystals [11] using the lattice electric field give sensitivities of $g_{a\gamma\gamma} \sim 10^{-9}$ GeV$^{-1}$. Geomagnetic Axion conversion [12] is a very interesting proposal for a satellite detector that will detect axion conversion on the earth magnetic field. Its sensitivity may reach $g_{a\gamma\gamma} \sim 10^{-11}$ GeV$^{-1}$. Helioscope techniques are going to be described in the next section for the CAST experiment. Tokyo Helioscope [13] and CAST [14] are currently active taking data. Microwave Cavity techniques will be described in the section for the ADMX [15] experiment. CARRACK [16] experiment uses Rydberg atoms to detect single photons from a microwave cavity where axions convert to photons.

### 3 The CAST experiment

In helioscope experiments the axion production hypothetically takes place in the core of the Sun where thermal photons interact with the nuclei to produce axions via the Primakoff conversion. The axion spectrum reflects the solar core temperature hence their energy spectrum has an average of 4.2 keV. The apparatus produces a very strong magnetic field where axion to photon conversion takes place [17] giving rise to X-rays that should be counted on top of the background. CAST uses a decommissioned LHC superconducting magnet operating at 1.8 K. It is 9.26m long, with a magnetic field B = 9 Tesla, equipped with two magnet bores of cross sectional area equal to 2 x 14.5 cm$^2$. The magnet lies on a rotating platform (Vertical: ±8°, Horizontal: ±40°) that is computer driven to track the sun during sunrise (1.5 hours) and sunset (1.5 hours). The rest of the time is devoted to background data recording. X-ray detectors equip all four bores of the magnet. The solar tracking precision of 1 arcmin is guaranteed by regular geometrical grid reference measurements and optical solar tracking during March and September every year. A Time Projection Chamber (TPC) covered both bores on the sunset side (Phase I) that was replaced by two Micromegas detectors during the second phase and on the sunrise side a Micromegas and a CCD detector connected to an X-ray focusing device. The focusing device – the ABRIXAS X-ray space telescope – helps to improve the signal to noise ratio by a factor of about 200. Figure 2 shows the CAST experiment.

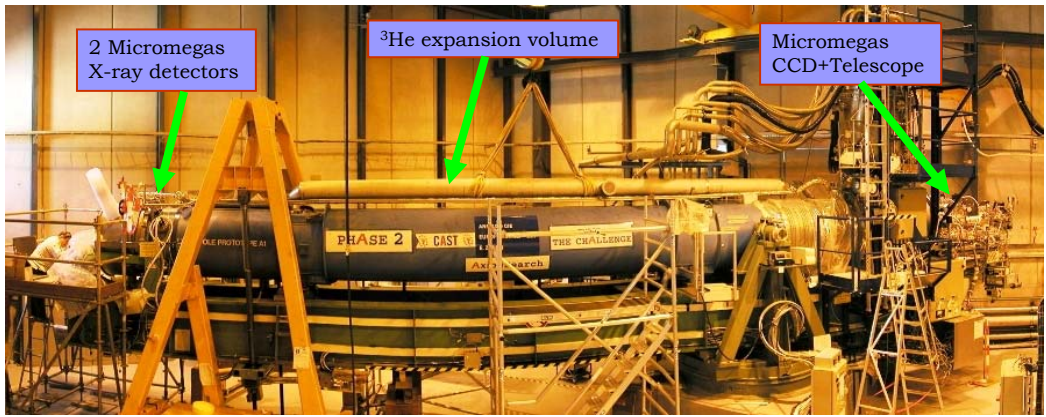

Figure 2. The CAST experiment. The magnet on the rotating platform, the X-ray detector and the ³He expansion volume are shown.

### 3.1 Operation in vacuum (Phase I)

The first phase of the experiment (Phase I) was completed during 2003 – 2004 with the magnet bores kept in vacuum. During Phase I CAST was sensitive to masses up to 0.02 eV. No excess signal was detected and the best experimental limit on $g_{\alpha\gamma\gamma}$ was set, surpassing for the first time the astrophysical limit from the Horizontal Branch stars evolution: $g_{a\gamma\gamma} < 8.8 \times 10^{-11} GeV^{-1} \, at \, 95\%CL$ [18] (Figure 3, left part). CAST has set the best experimental limit on $g_{\alpha\gamma\gamma}$ up to masses of 0.02 eV apart for the μeV mass range that is dominated by the microwave cavity experiments.

### 3.2 Operation with He (Phase II) – Visible tests - Prospects

Beyond $m_\alpha \sim 0.02$ eV the axion – photon coherence is lost in vacuum. To restore it and extend the sensitivity to higher masses a refractive gas is inserted in the magnet bore in consecutive steps. Each step provides the sensitivity only for a specific axion mass thus requiring several hundred steps to cover the mass range up to 0.39 eV with ⁴He as refractive gas. This part of Phase II was completed [19] in 2005 – 2006 and required 160 steps to reach a gas pressure of 13.4 mbar. To reach even higher masses a more sophisticated system was designed and built in order to use ³He as buffer gas. ³He remains in gas state for pressures up to 135.6 mbar allowing to explore masses up to 1.2eV. CAST has already scanned masses up to 0.65 eV with ³He but no excess events has been recorded. Figure 3 shows the exclusion plot for Phase II.

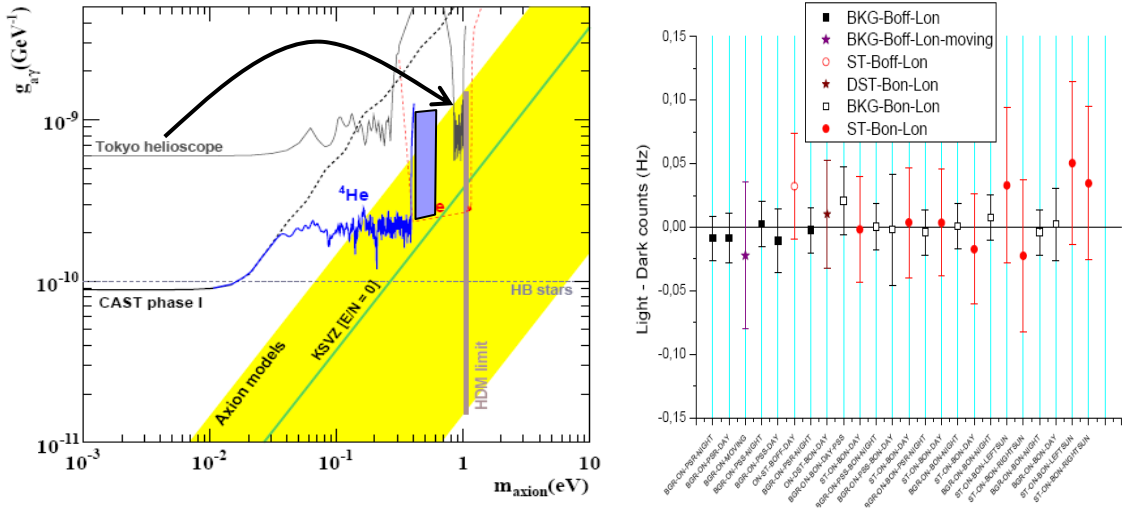

Figure 3. a)The CAST Phase II exclusion plot. The shaded (blue) area has been recently scanned and presents no axion signal. Recent Tokyo helioscope results are also shown near 1 eV. b) No excess of events were counted during the visible data taking.

The Tokyo Helioscope has scanned masses near 1 eV and has currently the best limit in that mass range [20]: $g_{a\gamma\gamma} < 5.6 - 13.4 \times 10^{-10} GeV^{-1}$ for $0.84 eV < m_a < 1 eV$

CAST was operated for the first time as an helioscope in the "visible" [21]. Optical devices (PMT and HPD) were coupled on the magnet bore and the Sun was tracked during a period of one week. No excess events were recorded (Figure 3b). This study can be used to set limits on paraphoton coupling.

CAST will explore the axion mass range beyond 1 eV and will close the gap with the upper axion mass limit provided by WMAP [22]. The new generation Micromegas detectors [23], built with the innovative micro-bulk technology, have shown extremely low background rates, about two orders of magnitude better than the micromegas currently used in CAST. The CCD – X ray telescope team is investing on improving their sensitivity to the sub keV range. This is of great interest to study phenomena like the Solar Corona heating mystery. Revisiting Phase I conditions (operation in vacuum) may offer the possibility to improve by almost an order of magnitude the exclusion limit.

### 4 The ADMX experiment

The ADMX experiment is a microwave cavity experiment and is based on the same idea proposed by P. Sikivie in 1983 [17] for the axion to photon conversion in a magnetic field. It aims at detecting relic axions which are either thermalized milky way axions or newly in falling axions with specific energy. The main elements of the experiment are a microwave cavity immersed in a strong magnetic field, a mechanism of tuning rods to scan a large range of frequencies (axion masses) and ultra sensitive amplifier linked to the readout chain. The readout chain is split to two: the Medium Resolution channel for the thermalized axions ($\Delta E/E \sim 10^{-6}$) and the High Resolution channel for coherent axion flows ($\Delta E/E \sim 10^{-22}$) [24]. The experiment operated using conventional heterojunction amplifiers (HFET) and recently they were replaced by SQUID (Superconducting, QUantum Interference Device) amplifiers that have a sensitivity of $\sim 10^{-26}$W, an order of magnitude more quiet than the GaAs HFET amplifiers. Figure 4 shows recent Medium Resolution limits in comparison with the recent SQUID preliminary results in Medium and High Resolution channels.

The current noise level that characterizes the ADMX sensitivity is $T_{sys} = T_{phys} + T_{ampl} = (1.3 + 0.4)$ K $= 1.7$ K. The effort is to minimize this noise by adding dilution refrigeration that will bring the total noise down to $T_{sys} = 200$mK.

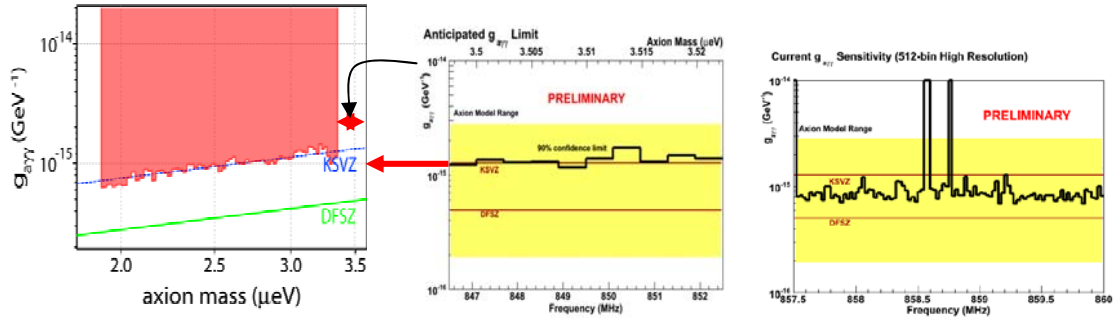

Figure 4. 1) Left: Medium resolution limits with the HFET amplifier 2) Middle: preliminary results for the Medium resolution search with SQUID and 3) Right: preliminary results for the High resolution search with SQUID.

## 5 Conclusions

The origin of Dark Matter and the Strong CP-problem are two long standing questions that remain yet unanswered. Axion discovery would certainly address the answer to both problems. Experimental effort has taken a boost the last years for the discovery of the elusive axion. Research on axions has on the other hand triggered large number of theoretical studies. Axion is an important constituent in string theory [25]. Very interesting ideas appear in recent studies like the one proposed by P. Sikivie on the Bose-Einstein Condensation of Dark Matter Axions [26]. The CAST experiment has explored a large area of the $g_{\alpha\gamma\gamma} - m_\alpha$ phase space and has set limits below the astrophysical constraints. Currently it is scanning masses approaching the WMAP upper axion mass limit of 1eV, inside the theoretically favored parameter space. New developments in the detector front may offer the possibility to explore a large fraction of the remaining phase space. In parallel with the X-ray program, CAST is active at the sub-keV range and the visible, offering the possibility to explore the Solar Corona heating mystery or study surface solar phenomena e.g. solar flares. ADMX on the other hand, using new SQUID amplifiers, are improving their sensitivity and they scan masses at the μeV range. If a new dilution refrigerator is used, both KSVZ and DFSZ models may be completely covered. This would enable in the future a definitive experimental search at the CDM interesting region.

**Acknowledgements**

We thank CERN for hosting the experiment. We acknowledge support from NSERC (Canada) , MSES (Croatia) grand No. 098-0982887-2872, CEA (France), BMBF (Germany) grant Nos. 05 CC2EEA/9 and 05 CC1RD1/0, the DFG grand HO 400/7-1, GSRT (Greece), RFFR (Russia), the Spanish Ministry of Science and Education (MEC) grands FPA2004-00973 and FPA2007-62833 and the Turkish Atomic Energy Authority. This work was partly supported by the U.S. Department of Energy and Contract No. DE-AC52-07NA27344; support from LLNL warmly acknowledged.